\documentclass[11pt,twocolumn]{article}
\begin{document}
\input{psfig}
%
  \advance\oddsidemargin  by -0.5in 
  \advance\evensidemargin by -1in
\marginparwidth          1.9cm
\marginparsep            0.4cm
\marginparpush           0.4cm
\topmargin               -0.2cm
\textheight             21.5cm
\hoffset +15mm
\newcommand{\be}{\begin{equation}}
\newcommand{\ee}{\end{equation}}
\newcommand{\bq}{\begin{eqnarray}}
\newcommand{\eq}{\end{eqnarray}}
\newcommand{\bibit}{\nineit}
\newcommand{\bibbf}{\ninebf}
\def\id{{\rm 1\kern-.21em 1}}
\font\nineit=cmti9
\font\ninebf=cmbx9
\title{\bf {Dynamics of twin-condensate configurations
in an open chain of three Bose-Einstein condensates} }
\author{P. Buonsante$^1$, 
R. Franzosi$^2$, and V. Penna$^1$ 
\\
\\
{\it $^1$Dipartimento di Fisica \& Unit\`a INFM, Politecnico di
Torino,}
\\
{Corso Duca degli Abruzzi 24, I-10129 Torino, Italy }
\\
{\it $^2$Dipartimento di Fisica \& Sezione INFN,
Universit\`a di Pisa, } \\ 
{Via Buonarroti 2, I-56127 Pisa, Italy.}
}
\begin{twocolumn}
\maketitle
\noindent
{\bf Abstract} --
The dynamics of a mean-field model of three coupled 
condensates is studied for initial states characterized 
by twin condensates. Condensates occupy an open 
three-well chain, where the central potential-well depth 
is an independent parameter.
Despite its diversity from the closed-chain symmetric model, 
such model is shown to have an integrable regime involving
a predictable evolution for twin-condensate initial states. 
After establishing
the dependence of the phase-space topology on the model parameters, 
the latter are shown to allow the control of
various macroscopic effects such as rapid population 
inversions, pulsed phase changes, self-trapping and
long-duration periodic oscillations.
%
Finally, based on quantizing the system
in terms of suitable ``microscopic'' canonical variables, the stability 
conditions of the integrable regime are related to the su(1,1) structure 
of the Hamiltonian spectrum.
\medskip

\noindent
{\bf Conference topic}: Bose-Eistein condensation of trapped atoms.
{\bf Program report number}: 6.3.1\\


\noindent
{\bf e-mail} address: penna@polito.it

\vfill\eject

\centerline{1. INTRODUCTION}
\medskip

The increasing skill in realizing experimental set-up's where
Bose-Einstein condensates (BEC) are distributed in potential 
wells with complex space structures~\cite{TRAP,Morsch} has prompted, 
over the last few years, the study of models such as the 
two-well system (dimer)~\cite{milb,RAG,aub,FPZ}, the three-well 
system (TWS)~\cite{hg,NHMM,FP2}, and their many-well generalization
such as the linear chain~\cite{CH} of coupled BECs. An aspect that has 
raised particular interest is the recognition of characteristic
behaviors and macroscopic phenomena that distinguish the 
dynamics of coupled wells.
The latter consists of interwell boson exchanges enacted by 
the tunnel effect~\cite{DALEG} that determine, in each well, 
macroscopic changes of the condensate phase and population.

The study of the TWS on a closed chain (three wells 
mutually coupled) has revealed, within a dynamics dominated 
by chaos~\cite{FP3}, the presence of an integrable 
subregime~\cite{NHMM,FP2} corresponding to (a phase-space submanifold of)
states in which two of the three wells exhibit identical condensate 
phases and boson numbers per well. Such twin-condensate (TC) configurations 
are preserved dynamically and make the TWS equivalent, in practice, to  
a two-well system. Particularly, the dimer integrability~\cite{FPZ} is 
inherited by TWS when this is prepared in an initial state with TCs. 

The TWS integrable subdynamics raises interest for various reasons. 
First, achieving a full knowledge of the TWS phenomenology with a 
regular dynamical character (this extends the analysis of 
Ref. \cite{BFP1}) is necessary to design 
significant experiments.
Second, since perturbing dimerlike states leads 
to prime chaotic behaviors, the modalities governing their 
onset (namely the preparation of suitable initial states
priming chaos) are a central aspect.
This is important, not only for the TWS, but also for the new 
class of systems (coupled-BEC chains) whose control, at the 
experimental level, is rapidly increasing.
Finally, in view of the last observation, the recognition 
of a class of states having regular time behaviors (near 
those showing chaos) is certainly useful to study the
manifestation of classical instabilities at the quantum 
level.

In this paper we consider the open TWS, namely a three-well
system in which two decoupled (lateral) wells interact with 
a third (central) one.
In particular, we show that, as in the closed-chain case, 
the open-TWS dynamics is equipped with an integrable subregime. 
To do this, we find new coordinates that both account for
the dynamical constraint embodying the TC form and 
preserve the canonical structure in the TC subdynamics 
without resorting to Dirac's procedure for constrained 
Hamiltonians.
As to the possibility to influence the system dynamical 
behaviour (and thus to play a relevant role in designing 
possible controlled experiments), we point out 
that in addition to the tunneling parameter (denoted by $T$)
the open TWS model exhibits a further adjustable parameter 
$w$ that distinguishes the central potential-well depth
from the two (equal) lateral-well depths. 

The present analysis prosecutes the work both of Ref. \cite{FP2} 
concerning the self-trapping onset in the dimerlike 
regime of the symmetric ($w=0$) closed TWS, 
and of Ref. \cite{BFP1}. In the latter the analysis of the
fixed points and the origin of chaotic behaviors in the open TWS 
were embodied, for arbitrary $T$ and $w$, in two dynamical phase diagrams
(relevant to the dimerlike and nondimerlike classes of fixed points)
that supply an exhaustive, operationally valuable, account of 
the system stability.

In the sequel, we concentrate on the dynamics of 
TC configurations and, more specifically, on the features 
``in large'' of orbit bundles that characterize the 
topology of the reduced phase space in the dimerlike
sub-regime. We investigate both the conditions under which
the system evolution exhibits either oscillatory or ballistic 
motion, and the possibility to control the self-trapping and 
other effects through the parameters $T$ and $w$.
Also, we show that the independence of parameter $w$, 
which introduces further asymmetry in the model-Hamiltonian 
structure but improves the possibility of fitting experimental 
conditions, does not affect the (partial) integrability of TWS 
and enriches noticeably the dynamical scenario of the system.
Finally, we evaluate the stability of TC
dynamics under quantum perturbations
by partially recovering the quantized form of the
TWS Hamiltonian. To this end the dynamical degrees 
of freedom that, within the new canonical coordinates
mentioned above, account for the deviation from dimeric 
regime, naturally undergo the quantization when they have a 
nonmacroscopic character. The stability of dimerlike 
configurations will be proven by relying on the su(1,1) 
structure exhibited by the model Hamiltonian.
\bigskip

\centerline{2. TRIMER-DYNAMICS CANONICAL FORM}
\medskip

The asymmetric TWS we consider here is described by the 
second-quantized Hamiltonian
$$
H= U \sum^3_{i=1}\, n^2_{i} -v (n_1 + n_3) - v_2 n_2 
$$
\be
\quad 
-{\frac{T}{2}} 
\left (a^{\dagger}_1 a_2 + a^{\dagger}_1 a_3  + h.c. \right ) ,
\label{TRIM}
\ee
which one derives from the many-body quantum theory
of BECs through a three-mode expansion of the condensate 
field operator \cite{NHMM}. Parameters $U$, $T$, $v$, 
account for the interatomic scattering, the tunneling
amplitude, and the sided-well potential depth,
respectively; the central-well depth has, in general, an
independent value $v_2=v+w \ne v$.
Another element of diversity from the symmetric-TWS case
is due to the absence of 
interwell coupling between the first and the third well. 
The operators $n_{i}\doteq  a^{\dagger}_{i} a_{i}$
count the bosons in the $i$th well ($N=\Sigma_i n_i$), 
while the destruction(creation) operators $a_{i}$ 
($a^{\dagger}_{i}$) obey the canonical commutators
$[a_{i},a^{\dagger}_{\ell}]= \delta_{i\ell}$.
Preceding studies of the TWS dynamics have been
focused on a case where the asymmetric character issues 
from nonconstant tunneling parameters $T_{ij}$ 
such that $T_{12} \gg T_{13}= T_{23}=T$.
Classically ($a^{\dagger}_{\ell} \equiv
a_{\ell}^{*}$, $a_{i} a^{\dagger}_{\ell}
= a^{\dagger}_{\ell} a_{i}$), the asymmetric TWS
has revealed~\cite{hg} the presence of homoclinic chaos,
while, quantally, the survival of breather
configurations has been investigated on the
trimer viewed as the smallest possible closed chain.
The Heisenberg equations related to $H$
for the boson operators $a_{i}$, $a^{\dagger}_{i}$
give, within the random-phase approximation, the
equations ($s = 1, 3$)
\bq
\begin{array}{c}
i\hbar {\dot z}_s = ( 2U|z_s|^2 -v ) z_s
- \frac{_{T}}{^2} z_2 \; , \cr
{\-} \cr
i\hbar {\dot z}_2 = [ 2U|z_2|^2  -(w+v )] z_2 
- \frac{_{T}}{^2} (z_3+z_1) \; , 
\end{array}
\label{EM3}
\eq
for the expectation values
$z_i= \langle  a_{i} \rangle$, $z_i^* = \langle a^{\dagger}_{i} \rangle$,
of the three wells.
These entail $\Sigma_i |z_i|^2$ as a conserved quantity
replacing the (conserved) total boson number $N$ ($[N, H]= 0$).
Eqs. (\ref{EM3}) can be also obtained from
$$
{\cal H}( Z, Z^*) \equiv  U  \sum^3_{j=1} |z_j|^4 
-v (|z_1|^2 + |z_3|^2 ) 
$$
$$
- (v+w )|z_2|^2
-\frac{_T}{^2} (z^{*}_2 z_{3} + z^{*}_2 z_{1} + c.c. )
$$
by using the standard canonical Poisson brackets
$\{z^*_k , z_j \}= i \delta_{kj}/\hbar$. The latter
and ${\cal H}( Z, Z^*)$ can be also derived variationally 
within the coherent-state approach developed in 
Refs.~\cite{AP,ZFG}.
\bigskip

\noindent
{\bf Sub-dynamics of Twin-Condensate configurations}.
Based on Eqs. (\ref{EM3}), imposing the constraint $z_1 = z_3$
leads to identify an integrable subregime of TWS dynamics
which consists of the orbit bundles lying on the sub-manifold
${\cal P} := \{(z_1, z_2, z_3): z_1 = z_3, z_2 \in \bf C \}$
in the original phase space $\cal M$. 
This can be proven by introducing the new canonical variables
$ \xi = (z_1 - z_3)/\sqrt{2}$, $z = (z_1 + z_3)/\sqrt{2}$
such that
$\{\xi^* , \xi \}= i/\hbar = \{z^* , z \}$, and
$\{z^* ,\xi \} = \{\xi^* ,z_2 \} = \{z^*_2 ,z \}= 0$.
One finds that
$$
\!\! H = 
\frac{_U}{^2} \left 
[ ( |z|^2 + |\xi|^2)^2 +(z \xi^* + z^* \xi)^2 
+ 2|z_2|^4 \right ] \quad \quad \quad
$$
\be
\quad
-v {\cal N} -w |z_2|^2
- \frac{_T}{^{\sqrt{2}}} (z_2 z^* + z_2^* z)
\label{H2}
\ee
(${\cal N} = \Sigma_i |z_i|^2 = |z_2|^2+|z|^2 + |\xi|^2 $)
giving, in turn, 
$$
i\hbar {\dot z} =U (|z|^2 + |\xi|^2)z -v z + U(z \xi^* + z^* \xi)\xi -
\frac{_T}{^{\sqrt{2}}} z_2 \, ,
$$
\be
i\hbar {\dot z}_2 =[2U|z_2|^2 -(v+w)] z_2 -\frac{_T}{^{\sqrt{2}}} z \, , 
\label{EMR}
\ee
$$
i\hbar {\dot \xi} =
U (|z|^2 + |\xi|^2)\xi
-v \xi + U(z \xi^* + z^* \xi)z \, .
$$
Setting $\xi = 0$ (that is $z_1 =z_3$) eliminates the 
third equation, while the remaining pair of coupled equations,
with the substitution $z = {\sqrt 2} z_1$,
\be
\cases{
&$ i\hbar {\dot z}_1 =(2U|z_1|^2 -v) z_1-\frac{T}{2} z_2 $\cr
&${\-}$ \cr
&$ i\hbar {\dot z}_2 =[2U|z_2|^2 -(v+w)]  z_2 -T z_1\; , $\cr}
\label{EM4}
\ee 
coincide with Eqs. (\ref{EM3})
under the restriction $z_1 = z_3$ to the sub-manifold $\cal P$.
The two costants of motion corresponding to the energy
$$
H_0 = U \left 
( \frac{_1}{^2} |z|^4 + |z_2|^4 \right )-v |z|^2 
$$
$$
-(v+w) |z_2|^2
- \frac{_T}{^{\sqrt{2}}} (z_2 z^* + z_2^* z)
$$
and the total boson number
$N= 2n_1 \, +\, n_2 = |z|^2 + |z_2|^2$ ($n_i \equiv |z_i|^2$)
make the dynamics issuing from Eqs. (\ref{EM4}) integrable.
Notice that, while $H_0$ still generates the correct
equations for $z$ and $z_2$ through ${\dot z} =\{ z, H_0\}$,
${\dot z}_2 =\{z_2 , H_0\}$, the same Hamiltonian written in terms
of $z_1$, $z_2$ no longer provides Eqs. (\ref{EM4}). The useful role 
of the variables $z$, $z_2$, $\xi$ is thus to supply a new canonical 
scheme where the restriction to $\cal P$ can be enacted directly on 
$H$ bypassing Dirac's scheme for constrained systems. 
The resulting Hamiltonian $H_0$ can be further cast in the 
pendulum-like version
$$
H_0 = 
C_0 + \frac{3U}{8}D^2 -\frac{f_{\nu}}{4} D 
- \frac{_T}{^{\sqrt 2}} \sqrt{N^2-D^2} \cos 2\theta \, ,
$$
in which the motion constant $N$ is used explicitly,
$C_0 = 3UN^2/8 -N(v+w/2)$, $f_{\nu}= UN (1-2\nu)$,
$\nu=w/UN$, and $\theta :=(\phi_2 - \phi)/2$ 
[defined through 
$z= |z| \exp (i\phi)$, $z_2= |z_2| \exp (i\phi_2)$] 
and $D := |z|^2 -|z_2|^2$ 
fulfill the brackets $\{ \theta, D \} = 1/\hbar$. The ensuing representation 
on the plane $(\theta, D)$ is the simplest possible way to describe 
the space $\cal P$ and its structure changes when $T$, $U$, 
$w$ are varied.
Such a $\theta-D$ picture involves two equations
\be
\hbar {\dot \theta} =  
\frac{3}{4}U D - \frac{f_{\nu}}{4}+ 
\frac{T D \cos 2\theta }{\sqrt{2} \sqrt{N^2-D^2}} 
\, ,
\label{EQMO1} 
\ee
\be
\hbar {\dot D} = - \sqrt{2} T  \sqrt{N^2-D^2}\, \sin 2\theta \, ,
\label{EQMO2}
\ee
only. The other pair of canonical variables $\psi = (\phi_2 + \phi)/2$, 
$N= |z|^2 +|z_2|^2$, in fact, do not participate in the dynamics: while 
${\dot N} = 0$, the angle $\psi$ plays an auxiliary role since 
${\dot \psi}$ can be proven to be completely determined by the 
evolution of $\theta$ and $D$ which, on the contrary, have 
$\psi$-independent equations.

\bigskip

\centerline{3. PHASE-SPACE STRUCTURE}
\centerline{AND PARAMETER DEPENDENCE}
\medskip

The TWS integrable subdynamics is reducible to a 
one-dimensional potential problem with an unique 
(phase-independent) $E$$N$-dependent equation 
${\dot D}^2= f(D; E,N)$ for $D$
by implementing standard quadrature procedures~\cite{FP2,PR}.
The latter lead to solve explicitly the motion equations
in terms of elliptic functions. 
Nevertheless, the analytic complexity of the one-dimensional potential 
makes rather difficult to establish the dependence on $\tau$ and $\nu$
of the $\cal P$ structure as well as of the dynamics therein represented. 
Then, since we wish to recognize the features that 
distinguish the dynamical behaviors of TC configurations 
[rather than the exact analytic solutions of Eqs. (\ref{EQMO1}),
(\ref{EQMO2})], we simply approach the problem by performing in parallel 
the study of the fixed points of Eqs. (\ref{EQMO1}), (\ref{EQMO2}) 
and the numerical reconstruction of the phase-space portraits.
This furnishes in a quite direct way information on both the topology 
changes in $\cal P$ when $\tau$ and $\nu$ are varied, and the features 
``in large'' of the dimerlike bundles of trajectories represented on 
the $\theta-D$ plane.
In particular, the conditions that cause variuos macroscopic effects 
in the dynamical behavior can be evidenced effectively.

The equation pair that determine the fixed points, 
$0 =[3UD - f_{\nu} +2\sqrt{2}T D \cos(2\theta)/r(D)]/4$,
$0=\sqrt{2} T r(D)\, \sin (2\theta)$
with $r(D)=\sqrt{N^2-D^2}$,
is stemmed from equations (\ref{EQMO1}), (\ref{EQMO2}) 
when setting ${\dot \theta} = 0 = {\dot D}$.

These supply maxima, saddles and minima that constitute 
the fixed-points set in the reduced phase space ${\cal P}$. 
The minimum is situated in $\theta =0$, $D \ne 0$ ($D=0$ 
only for $\nu=1/2$), while the other extremal points stay 
on the straight line $\theta = \pi/2$. 
Under such conditions, the values of $D$ at fixed-points are obtained from
the nonlinear equation
\be
0=\frac{3}{4}X - \frac{1-2\nu}{4}
+ \,  \frac{\sigma \tau X }{\sqrt{2} \sqrt{1-X^2}}
\label{SM1}
\ee 
where $\sigma = +1$ ($\sigma = -1$) in the case of the minimum
(maxima/saddles), and the adimensional quantities $\tau := T/UN$,
$X = D/N$ have been introduced in addition to $\nu := w/UN$. 
The number of fixed point as a function of $\nu$, $\tau$  
is easily deduced from eq. (\ref{SM1}). In the parameter space 
$(\nu, \tau)$, the critical lines
$$
\nu_r =  \frac{1}{2} +(-)^r\,
\frac{3}{2} \left [ 1 -2 \left ( \frac{\tau }{3} \right )^{2/3}
 \right ]^{3/2}\, ,
$$
($r = 1, 2$) are recognized to play a crucial role.
For $\nu_1(\tau) < \nu < \nu_2 (\tau)$ ($\nu = \nu_1(\tau), \nu_2(\tau)$) 
three (two) solutions are furnished by equation (\ref{SM1}) with $\sigma =-1$,
in addition to the permanent single solution of the case $\sigma =+1$.
The lobe 
$$
{\cal L} = \{ (\nu, \tau) : 0< \tau< 3/ {\sqrt 8},\,\,
\nu_1 (\tau) < \nu < \nu_2 (\tau) \},
$$
defined in the parameter plane $(\nu,\tau)$ identifies
completely the 
area corresponding to the three-solution regime. Outside ${\cal L}$,
only one solution survives together with the solution of the case 
$\sigma =+1$.
\bigskip

\centerline{4. TOPOLOGY CHANGES AND}
\centerline{DYNAMICAL EFFECTS}
\medskip

The recognition of the critical lines 
bounding $\cal L$ together with some numeric simulations of 
the reduced phase space $\cal P$ (see the figure pairs 1 and 2), 
enable us to determine the influence of $\tau$ and $\nu$ on the 
topology of $\cal P$, namely on the structure of orbit bundles 
of the TC dynamics.

For $(\nu,\tau) \in \cal L$,
the fixed points resulting from Eqs. (\ref{SM1}),
with $\sigma =+1$, consist of two maxima $M_-$, $M_+$ and 
a saddle $M_0$ such that $\theta = \pi/2$ and $X_- < X_0 < X_+$. 
Such a situation~\cite{com} is described in Fig. 1a and Fig. 1b where, with 
$\tau= 0.5$, the cases $\nu = 0.25$ and $\nu = 0.75$, respectively, 
are depicted. Their symmetric structure under
the global reflection with respect to the axis $D/N =0$,
reflects the symmetry of Hamiltonian $H_0$ which combines the 
changes $\nu \to 1-\nu$ and $D \to -D$.

When $\tau = const$ and $\nu \to \nu_1^{-}$ ($\nu \to \nu_2^{+}$), 
a {\it coalescence effect} takes place in which $M_-$ ($M_+$) and $M_0$ 
merge in a unique point. 
This process can be driven either by the potential-depth gap 
$w=v_2 -v$ or by the total boson number $N$ through the parameter 
$\nu=w/UN$ (notice that $U$, accounting for the s-wave scattering 
length, in the present paper is considered as a fixed parameter).
The presence of the adjustable gap $w$, however, allows one
to operate leaving $N$ (and thus $\tau = T/UN$) unchanged. 
Fig. 2a and Fig. 1b illustrate the effect of 
crossing the $\cal L$ border by changing $\nu$ from 
$1$ to $0.75$. No change of the $\cal P$ structure 
occurs by further decreasing $\nu$ as far as $\nu_1 (\tau) <\nu $.
Fig. 1a and Fig. 1b, in fact, have the same number of extremal points.

Thanks to the presence of $\tau$, the coalescence can be achieved 
as well if, keeping $\nu$ constant within the interval $-1 < \nu < 2$,
the change $\tau \to \tau'> \tau$ is such that 
$(\nu,\tau) \in {\cal L}$ while $(\nu,\tau') \not \in {\cal L}$.
The comparison of Fig. 1a and Fig. 2b, both having $\nu = 0.25$, 
well illustrates the phase-space structure emerging from the coalescence of 
$M_+$ with the saddle $M_0$ as a consequence of the change 
$\tau = 0.5 \to \tau' =  1$. 

The {\it orbit-bundle} structure deeply reflects the action of
$\nu$-$\tau$ changes.  
For $(\nu,\tau) \in  \cal L$, $\cal P$ displays $\it five$ independent
bundles whose topology is characterized by means of three 
separatices (dashed lines in all the figures) ${\cal S}_+$, ${\cal S}_-$, 
and ${\cal S}_0$ (the latter shows a $\gamma$ shape).
Outside $\cal L$, owing to the coalescence (see Figs. 2a, 2b),
separatrix ${\cal S}_0$ 
disappears, while the orbit bundles reduce to $\it three$.
Curves ${\cal S}_{\pm}$ are easily identified by observing that they 
start/terminate at $D= \pm N$, $\theta= \pi/4,\, 3\pi/4$, whereas
${\cal S}_0$ is based at $D= N X_0$, $\theta= \pi/2$, where $X_0$
is given by Eqs. (\ref{SM1}).
Notice that for $\nu <0.5$
the two basins encircled by ${\cal S}_+$ and ${\cal S}_-$ 
contain the minimum $\rm m$ and the maximum $M_-$, respectively 
(see Figs. 1a, 2b), while
the opposite situation appears for $\nu > 0.5$ (see Figs. 1b, 2a), 
where ${\cal S}_+$ and  ${\cal S}_-$
encircles $M_+$ and $\rm m$, respectively.
In both cases, such basins are filled by closed-curve bundles
entailing $\it bounded$ oscillations for $\theta$.

\noindent
Two bundles of {\it ballistic} orbits, instead, are comprised between 
${\cal S}_+$ and ${\cal S}_-$ and separated by
${\cal S}_0$ when $(\nu,\tau) \in  \cal L$.
In this case, no constraint limits $\theta$ that 
covers its whole range $[0,2 \pi]$, as shown by Figs. 3a, and 4a. 
The closed arc ${\cal C}_0$, constituing the noose of ${\cal S}_0$, 
confines a further closed-orbit bundle that encircles $M_+$ ($M_-$) 
for $\nu < 0.5$ ($\nu > 0.5$).

When the coalescence is enacted, ${\cal C}_0$ and the basins 
around $M_{\pm}$ collapse, while ${\cal S}_0$ merges to the 
ballistic-curve flows lying between ${\cal S}_-$ and ${\cal S}_+$ 
(see Figs 2a, 2b). 
This topology change is accompanied by another, dynamically relevant,
effect that consists of the suppression of the two motion curves
forming ${\cal S}_0$: the closed branch ${\cal C}_0$ and the 
$\Lambda$-shaped branch. Such orbits, both starting and 
terminating at $M_0$, are characterized by an $\it infinite$ covering 
time (illustrated in the following sub-section) as required by the saddle 
feature of $M_0$. Orbits placed in their proximity, which 
before the coalescence display 
arbitrarily long percurrence times, lose such a character after
the saddle suppression.
A special macroscopic trait therefore distinguishing
the parameter choices $(\nu,\tau) \in {\cal L}$ and 
$(\nu,\tau) \not \in {\cal L}$ is that the former choice
always involves the presence of nonzero set of initial 
conditions around $M_0$ whose characteristic time period 
can become arbitrarily large. This feature, of course, is interesting 
both for achieving a complete control of the TWS characteristic times
and in view of the experimental observations.
\bigskip

\noindent
{\bf Ballistic motion and self-trapping}.
In Figs. 3, where the elapsing time is $UNt$,
we illustrate the evolution of both $D/N$ (its
oscillations are evidenced in the grey stripe) and 
$\theta$ relevant to various orbits in $\cal P$.
In Fig. 3a, $D/N$ and $\theta$ are given for two orbits 
having energy values $E_A$, $E_B$, with $\tau = 0.5$, $\nu= 0.25$. 
These energies are such that 
$E_A-E_B \simeq 5 \cdot 10^{-4} E_0$, where $E_0$ is the energy
of ${\cal S}_0$ and $|E_B - E_0|\simeq  10^{-9} E_0$. The period
of the $E_B$ orbit is greater than that of the $E_A$ orbit since 
the former is closer to ${\cal S}_0$. Both $\theta$ and $D/N$
exhibit an intermittent evolution where time changes are
confined in a sub-interval of the characteristic period in which
rapid {\it population\, inversion} effects are visible as well.
Fig. 3b shows $D/N$ and $\theta$ for two orbits near 
$m$, and $M_-$ and allows one to compare the corresponding 
periods with the ballistic motions of Fig. 3a.
The evolution of $\theta$ and $D/N$ is displayed in Fig. 4a 
for an orbit almost coinciding
with ${\cal S}_0$ (rapid local variations of both are visible in 
correspondence to the motion around the ${\cal S}_0$ noose). 
Fig. 4b shows $\theta$ and $D/N$ for two orbits near ${\cal S}_-$
for $\tau= 1$, $\nu= 0.25$: the inner orbit exhibits a pulsed
population inversion where abrupt changes alternate with an
almost linear decreasing of $D/N$.

In consequence of the two-parameter dependence of the
$\cal P$ topology, also the {\it self-trapping} effect can be 
induced/suppressed through $\tau$ and $\nu$. We recall that
the latter is particularly evident~\cite{aub,FPZ} 
when the phase space displays pairs of isoenergetic orbits. 
In this case the system is imposed to single out one of the available 
orbits at a given energy. This symmetry breaking effect entails 
the self-trapping in a restricted region of $\cal P$ which 
cannot be left unless sufficiently large energy changes
are effected.

In the presence of two, well separated maxima $M_-$ and
$M_+$ (Fig. 1a and Fig. 1b), the self-trapping regimes
exhibit oscillations of populations $n_i= |z_i|^2$
such that either $n_1=n_3 \simeq 0$, 
$n_2 \simeq N$ (orbits close to $M_-$ with $D \simeq -N$)
or $n_1=n_3 \simeq N/2$, $n_2 \simeq 0$ (orbits close to $M_+$
with $D \simeq N$).
As a result, the trajectories encircling the maximum manifest 
an evident, stable population imbalance (namely $n_1= n_3 >\! >n_2$,
$n_1= n_3 <\!< n_2$) during the system evolution that excludes
the oscillations to approach intermediate states
$n_1=n_3 \simeq N/4$, $n_2 \simeq N/2$ ($D\simeq 0$)
near the minimum energy configuration $m$.
We wish to notice that, when the suppression happens and
$M_-$ ($M_+$) and $M_0$ merge in a unique point, in general, 
the remaining maximum $M_+$ ($M_-$) keeps a position that is 
rather different from that of the minimum
(see Fig. 2a, Fig. 2b) and thus preserves a regime
with a marked population imbalance.

Concerning the change $\nu \to 1-\nu$ 
($ \Leftrightarrow \,\, D \to -D$) 
it is worth noting that, while the basins of 
$M_0$ and $M_{\pm}$ remain unaltered, the ballistic orbit 
bundles between ${\cal S}_+$ and ${\cal S}_-$ undergo an 
abrupt inversion. For $\nu = 0.5$ the linear D term in $H_0$
disappears and the resulting symmetry $D\to -D$
renders the dimeric regime of the open TWS
equivalent to the pure symmetric dimer~\cite{RAG,FPZ}.
We devote a final remark to the case $\nu =0$ (equal well depths). 
Since $(\tau, \nu)\not \in \cal L$ for
any $\tau \ne 0$ if $\nu =0$, then $\cal P$ portraits are quite 
similar to the one of Fig. 2b showing the ballistic 
phase evolution and pulsed population inversions (see Fig. 4b).
It is worth noting that the evident 
asymmetric character of the TWS dynamics, 
well displayed by the $\cal P$ portraits, ensues not only 
from the well-depth diversity but also from the fact 
the pair of TC is described within the integrable regime
by an unique collective variable $z$.
\end{twocolumn}
\begin{onecolumn}
\begin{figure}
\centerline{\psfig{height=7cm,file=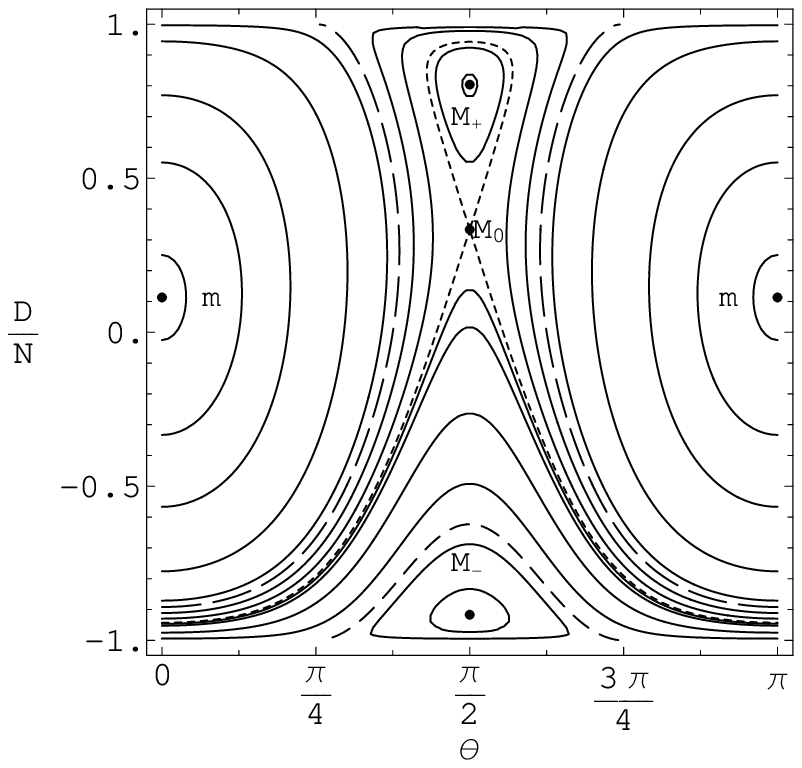}
\psfig{height=7cm,file=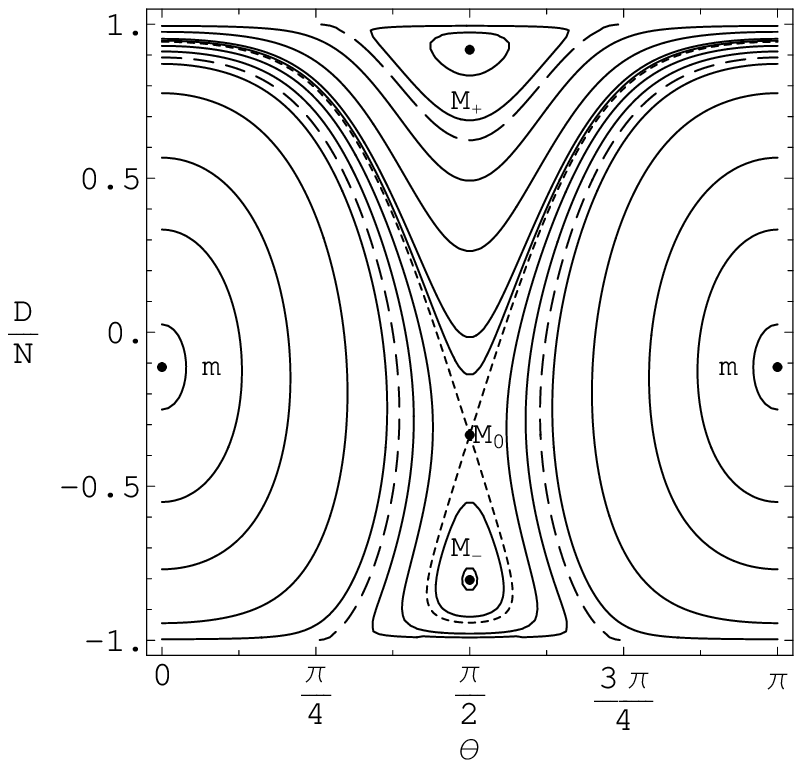}}
\caption{(a) Left figure is obtained by setting $\tau = 0.5$, $\nu= 0.25$;
(b) right figure corresponds to $\tau = 0.5$, $\nu= 0.75$. In both figures,
the minimum $m$, the saddle $M_0$ and the maxima $M_{\pm}$ are visible
as well as separatrices ${\cal S}_0$, ${\cal S}_{\pm}$ represented by 
different dashing styles.
}
\label{fig1}
\end{figure}
\begin{figure}
\centerline{\psfig{height=7cm,file=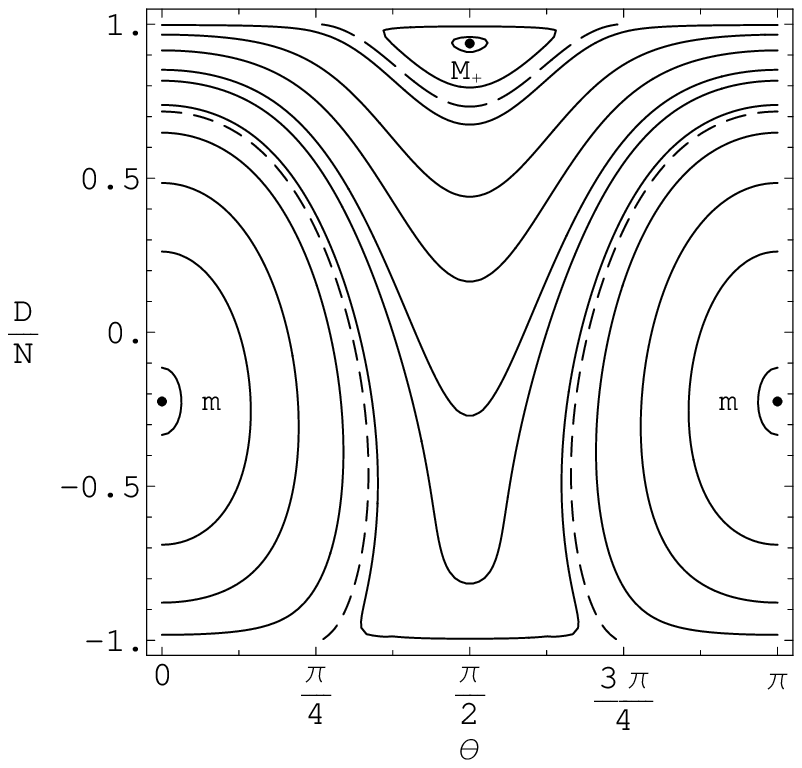}
\psfig{height=7cm,file=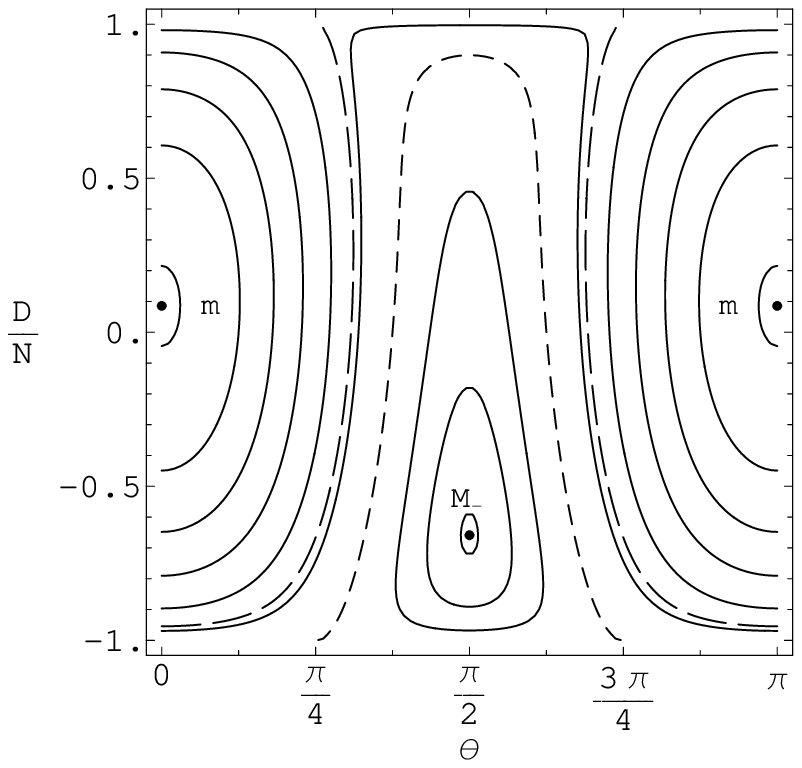}}
\caption{(a) Left figure illustrates $\cal P$ for $\tau = 0.5$, $\nu= 1$:
increasing $\nu$ from $\nu= 0.75$ causes the coalescence of $M_0$ and 
$M_{-}$ displayed in Fig. 1b.
(b) Right figure corresponds to $\tau = 1$, $\nu= 0.25$: increasing $\tau$ 
from the value $\tau = 0.5$ causes the coalescence of $M_0$ and $M_{+}$ 
displayed in Fig. 1a.
}
\label{fig2}
\end{figure}
\end{onecolumn}
\begin{twocolumn}
\noindent
\newpage
\end{twocolumn}
\newpage
\begin{onecolumn}
\begin{figure}
\centerline{\psfig{height=7cm,file=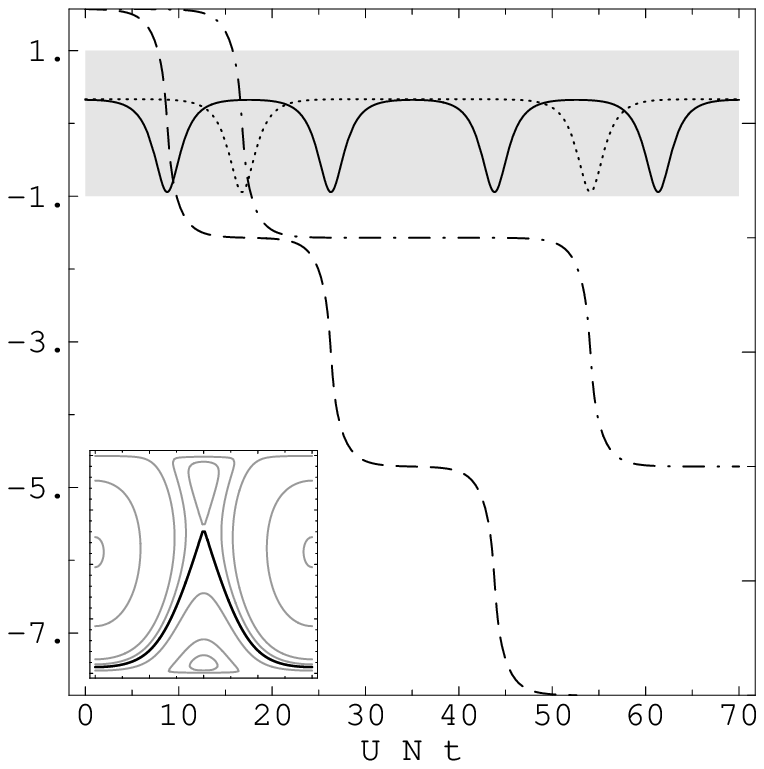}
\psfig{height=7cm,file=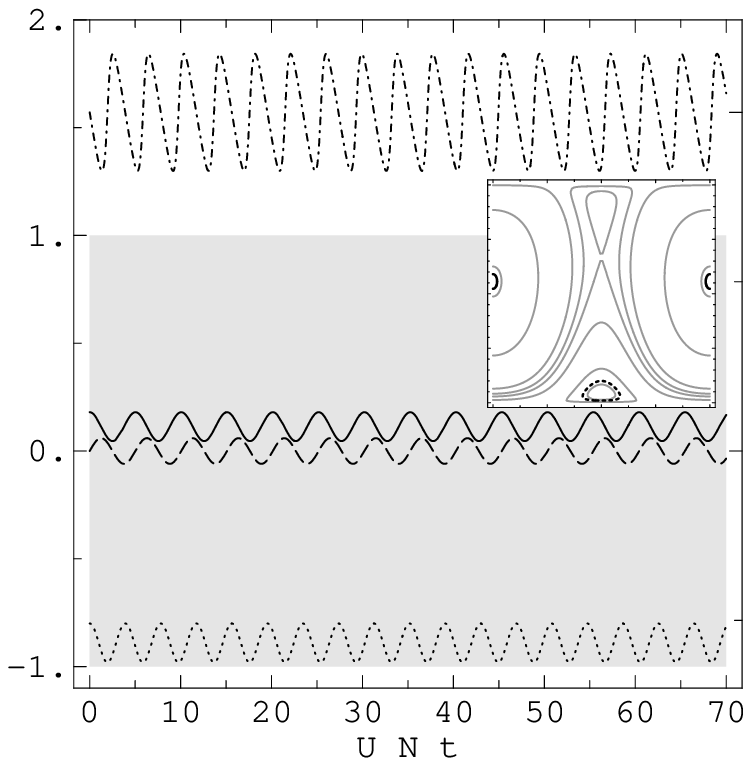}}
\caption{Evolutions of $D/N\in [-1,1]$ (in the grey stripes) and
$\theta$ for some orbits in $\cal P$ (see the sub-panels), 
with $\tau =0.5$, $\nu= 0.25$; $UNt$ is the rescaled time.
(a) Left figure: $\theta$ and $D/N$ for two orbits very 
close to the $\Lambda$ branch of ${\cal S}_0$.  
The orbit closer to ${\cal S}_0$ has a larger period. 
(b) Right figure: periods of left-figure orbits can be compared
with those of two orbits near $M_-$ and $m$.
Refer to the text for further comments.
}
\label{fig3}
\end{figure}
\begin{figure}
\centerline{\psfig{height=7cm,file=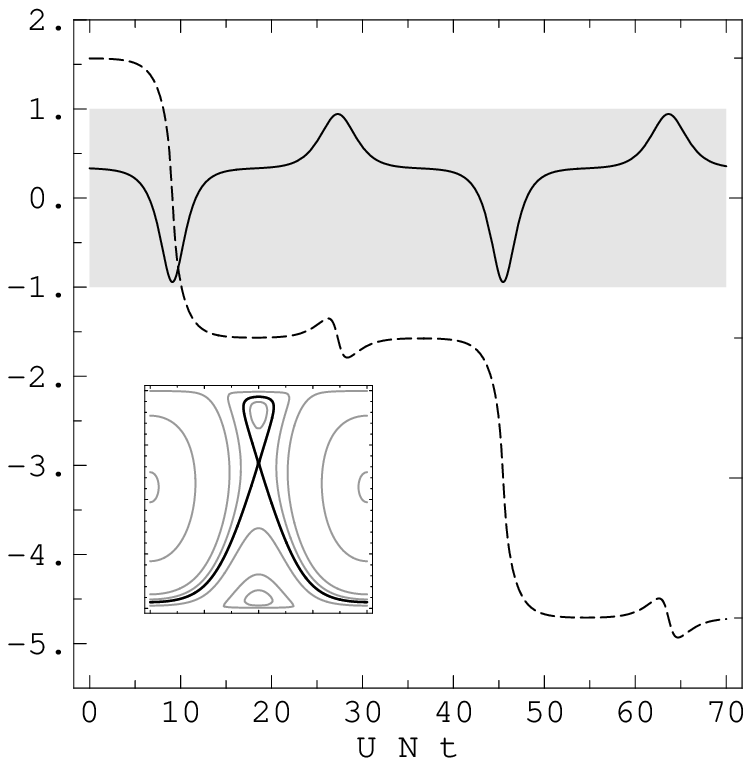}
\psfig{height=7cm,file=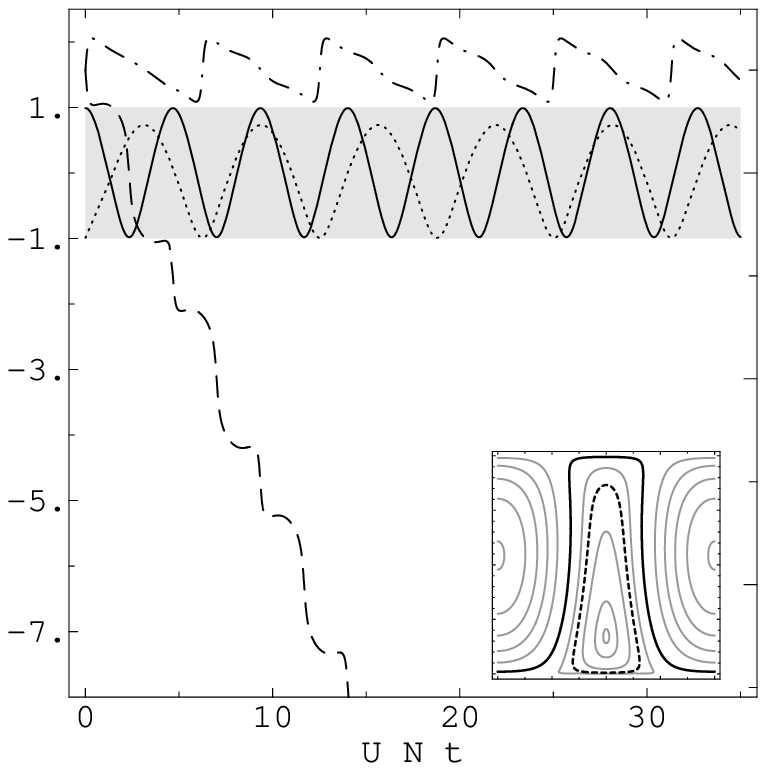}}
\caption{
Evolutions of $D/N \in [-1,1]$ (evidenced in the grey stripes) and $\theta$
for various orbits in $\cal P$ (see the sub-panels). $UNt$ is the rescaled time
(a) Left figure corresponds to an orbit very near ${\cal S}_0$, with 
$\tau = 0.5$, $\nu= 0.25$.
(b) Right figure shows two orbits adjacent to ${\cal S}_-$ 
with $\tau = 1$, $\nu= 0.25$.
}
\label{fig4}
\end{figure}
\end{onecolumn}
\begin{twocolumn}

\centerline{5. QUANTUM APPROACH TO}
\centerline{ DIMERLIKE DYNAMICS}
\medskip

The trimer dynamics naturally undergoes
a quantum formulation both when it is investigated
in the low-energy regime and if small particle numbers per well
do not allow Bogolubov's approximation. 
Classically (namely in the mean-field picture described in section 2), 
the ground-state configuration is
fully determined and represents a special case within the dimeric regime.
In the alternative canonical scheme discussed above, quantum Hamiltonian
(\ref{TRIM}) 
takes the form $( n_c := c^{\dagger}c, \, c= a, b)$
$$
H= U \left [ n_2^2 + \frac{_1}{^2}( n_a + n_{b} )^2  +
\frac{_1}{^2}(b^{\dagger} a + a^{\dagger} b )^2 \right ]
$$
$$
-v (n_a + n_{b}) - w n_2 -
{\frac{T}{\sqrt 2}}\, \left (a_2 a^{\dagger}  + a_2^{\dagger} a  \right ) ,
$$ 
where the quantum counterparts of new variables $\xi$, $z$ are
$$
b = (a_1 - a_3)/\sqrt{2} , \quad
a = (a_1 + a_3)/\sqrt{2}\, ,
$$
obeying the commutators
$[b, b^{\dagger}] = 1 = [a , a^{\dagger} ]$. 
In the spirit of Bogolubov's approximation (which is valid when a bosonic 
mode $c$ is strongly occupied so that $c^{\dagger} \simeq c^*$ and 
$c^{\dagger} c \simeq |c|^2$ with $[c, c^{\dagger}] \simeq 0$ ),
we observe that for states close to the dimeric regime
both $a$ and $a_2$ are macroscopic, while $b$, $b^{\dagger}$
can be seen as microscopic degrees of freedom describing the deviation 
from the pure dimeric states. In particular, the classical 
condition $\xi \simeq 0$ should correspond quantally to have 
the expectation value $\langle b \rangle \simeq 0 $, whereas 
$|\xi|^2$ -classically, the population of the $\xi$ mode- should
correspond to the operator $n_{b}$. Only quantum states involving
small quantum numbers for $n_{b}$ are expected to be 
excited in the system dynamics. 

In the light of this,
Eqs. (\ref{EMR}) can be employed to approach in
the appropriate way the dynamics
in which $b$, and $b^{\dagger}$ are quantum objects, 
whereas $a$ and $a_2$ 
can be identified with $z$ and $z_2$, respectively, 
owing to their macroscopic 
character. The first two equations are in fact decoupled from the third
one if one considers only the first order terms in $\xi$, $\xi^*$.
Under such assumption the third equation, where $|\xi|^2 \xi$ is now 
negligible, is independent in that the time evolution of $z$-dependent 
parameters is determined by the first two equations. Also, its form
implies that $|\dot \xi| << |\dot z|, \, |{\dot z}_2|$ thus making $\xi$
a slowly varying variable.   
This scenario suggests that the quantum problem for the $\xi$ mode 
can be described by the Hamiltonian
$$
H_b =  \frac{_U}{^2}( 2|z|^2 n_{b} + n^2_{b} )  +
\frac{_U}{^2}(b^{\dagger} z + z^* b )^2 -v n_{b} \, ,
$$ 
where $n^2_b$ is negligible with respect to $2|z|^2 n_{b}$. The spectrum
of $H_b$ is easily calculated by recognizing that its spectrum generating 
algebra~\cite{PE96} is su(1,1). The latter is generated by
$$
K_3 =(2 n_b +1)/4  ,\,\,
K_+ = (b^{\dagger})^2/2  
,\,\, K_- = (K_+)^{\dagger},
$$
with commutators
$[K_3, K_{\pm}] =$ $\pm K_{\pm}$, $[K_{-}, K_{+}]$ $= 2 K_3$.
Upon noticing that 
$(b^{\dagger} z + z^* b )^2 =$ 
$ z^2 (b^{\dagger})^2  +$ $ (z^*)^2 b^2 +$
$|z|^2 (2 n_{b} +1)$, 
one gets $H_b$ in the new form
$$
H_b = 2 (2U|z|^2 -v)K_3 -\frac{_U}{^2} |z|^2 
$$
$$
+U[ z^2 K_+ + (z^*)^2 K_-]  +\frac{v}{2}\, .
$$
By using the unitary transformation $U_3 =$ $\exp [i\phi K_3]$,
(notice that 
$U_3 K_{\pm} U_3^{\dagger}$ $= e^{\pm i\phi}  K_{\pm}$),
where $\phi$ is given by $z = |z| e^{i\phi}$, one finds that
$H_b =  U_3 H_{*} U_3^{\dagger}$ with
$$
H_{*} = 
%
2 (2U|z|^2 -v)K_3 + 2U|z|^2 K_1 -\frac{_U}{^2} |z|^2  +\frac{v}{2} 
\, ,
$$ 
where $K_1 = (K_+ + K_-)/2$ is one of the two noncompact generators 
of the pseudo-angular momentum standardly defined by means of
$K_{\pm} = K_1 \pm i K_2$. The further transformation
$U_2 = \exp [-i\alpha K_2]$ such that
$$
U_2 K_3 U_2^{\dagger} = {\rm ch}(\alpha)  K_3 + {\rm sh} (\alpha)  K_1
\, ,
$$
allows $H_{*}$ to be expressed as
$$
H_{*} =  U_2 \left ( R\, K_3 + 
-\frac{_U}{^2} |z|^2  +\frac{v}{2} \right )U_2^{\dagger} 
$$
$$
=
R\, {\rm ch} (\alpha)  K_3 + R\, {\rm sh} (\alpha)  K_1
-\frac{_U}{^2} |z|^2  +\frac{v}{2}
$$
provided ${\rm th} \alpha = U|z|^2/(2U|z|^2 -v)$
and $R^2 = (3U|z|^2 -v)(U|z|^2 -v)$, 
and either $|z|^2 >v/U$ or $|z|^2 <v/3U$ is fulfilled.
Whenever one of the latter conditions is violated 
($v/3U \le |z|^2 \le v/U$) the transformation
$U_2$ must be related to the noncompact element $K_1$
(instead of $K_3$) in order to reproduce $H_{*}$.
The ensuing effect is dramatic: While $K_3$ has a discrete, 
positive spectrum bounded from below, 
$$
spect(K_3) = (2 n +1)/4 \, , \quad n = 0, 1, 2,...,
$$
$K_1$ has a continuous,
unbounded spectrum. Since we have used unitary transformations
such spectra are those characterizing $H_b$ itself.  
Therefore, only in the case when the initial conditions
on $z$ ensure the presence of a discrete spectrum
$$
E_n(|z|^2 ) = 
\sqrt{(3U|z|^2 -v)(U|z|^2 -v)}\,\frac{(2n+1)}{4} 
$$
$$
-\frac{_U}{^2} |z|^2  +\frac{v}{2}
$$
the ground-state turns out to be stable in that an energy gap
separates it from the first excited state. This makes stable the 
dimeric regime.
A final check is quite natural at this point.
Making use of the first of eqs. (\ref{EMR}) in the
integrable sub-dynamics ($\xi=0$)
and recalling that $z=\sqrt 2\,z_1$, the discreteness condition 
becomes 
\begin{equation}
\label{E:disc1}
\left(4U \,|z_1|^2+\frac{T}{2}\frac{z_2}{z_1}\right) 
\frac{T}{2}\frac{z_2}{z_1}\geq 0
\label{discon}
\end{equation} 
Remarkably,  the conditions on the Hamiltonian parameters 
$\tau$ and $\nu$ enforcing condition (\ref{E:disc1}) are basically 
the same ensuring that the classical fixed points are stable. That is 
to say, the discreteness condition (\ref{discon})
selects the stable regions (maxima and minima) of the dimeric stability 
diagram appearing in ref. \cite{BFP1}.

We briefly recall that, owing to a global phase symmetry, each dimeric 
fixed point can be identified by a pair of real numbers, $x_1$ and $x_2$, 
such that $z_1=z_3=\sqrt N\,x_1$ and $z_2=\sqrt N\,x_2$. It proves then 
useful to introduce the real parameter $\alpha=x_2/(\sqrt 2 x_1)$ such 
that $x_1 = \cos(\vartheta)/\sqrt 2$, $x_2 = \sin(\vartheta)$, where 
$\vartheta=\arctan \alpha$. This way each fixed point is uniquely related 
to a single real parameter, either $\vartheta$ or $\alpha = \tan \vartheta$. 
It turns out that, for a given choice $\tilde\tau$,$\tilde\nu$ of the 
parameters, the relevant $\vartheta$'s can be found  by intersecting the 
straight line $\tau=\tilde\tau$ and the curve 
$\tau=\max[0,\tau_{\tilde\nu}(\vartheta)]$, 
where $\tau_{\nu}(\vartheta)=[\nu+1+\alpha^2(\nu-2)]/(\alpha^4-1)$.
If $\vartheta>0$ the relevant fixed point is a minimum. Conversely, if  
$\vartheta\leq 0$, it is either a maximum or an unstable \cite{COM1}
saddle depending on 
whether $\tilde \tau>\bar \tau(\vartheta)$ or 
$0<\tilde \tau<\bar \tau(\vartheta)$, where
$\bar \tau(\vartheta)=\max[\tau_1(\vartheta),\tau_2(\vartheta)]$,
%
%
\be
\tau_1(\theta)=\frac{-2\sqrt 2}{\beta\,(1+\beta^2)},\,\,
\tau_2(\theta)=
\frac{-6\sqrt 2 \beta^3}{(1+\beta^2)^3}, 
\ee
with $\beta= \tan \theta$.
Now, recalling that $T= U N \tau$, equation (\ref{E:disc1}) finally
reduces to 
%
\be
[2 \tau  \tan \theta /(1+\tan^2 \theta)+
\tau^2 \tan^2 \theta /\sqrt{2} ]\geq 0
\label{E:disc3}
\ee
which selects the minima and, almost exactly, the maxima regions 
defined by the study of the stability character of the classical 
fixed points above recalled. Indeed the points in the minima region, 
$\{0<\vartheta<\pi/2,\tau \geq 0\}$, trivially meet condition 
(\ref{E:disc3}). 
The same is true for most of the maxima region, 
$\{-\pi/2<\vartheta<0,\tau > \bar \tau(\vartheta)\}$. Indeed, 
if $\vartheta<0$, condition (\ref{E:disc3}) selects the region 
above $\tau_1(\vartheta)$, which is exactly the borderline between 
maxima and saddles when $\tau_1(\theta)>\tau_2(\theta)$. In the 
complementary case, $\tau_1(\theta)<\tau_2(\theta)$, a small portion 
of the saddles region is predicted to give rise to a discrete rather 
than continuous spectrum. This slight disagreement can 
be tentatively explained by observing that in the latter case 
$\vartheta$ (is likely to be) can be quite close to $-\pi/2$, 
which ultimately means $|z_1|\sim 0$. This, however, is in 
contrast with our  hypothesis that $z_1$ is macroscopic.
\bigskip

\centerline{5. CONCLUSIONS}
\medskip

Based on a variational technique previously developed~\cite{FPZ,AP},
we have studied the integrable sub-regime of the TWS dynamics in terms 
of the expectation value $z_i$ ($z^*_i$) relevant to operators $a_i$ 
($a^{\dagger}_i$). Our approach is equivalent to the standard 
mean-field picture.
Upon defining suitable collective canonical modes,
in section 2 we have presented an alternative description of the 
dynamics which allows to project the TWS dynamics on the reduced 
($z_1 = z_3$) space $\cal P$ containing the integrable orbits,
and avoids the nontrivial Dirac's method for constrained hamiltonian 
systems. The new picture thus obtained, in which TCs
are always present, turns out to
be equivalent to an asymmetric dimerlike model whose integrability 
is well known.
  
After deriving the fixed points of $H_0$ and the conditions 
determining the coalescence of fixed-point pairs depending 
on $\tau$ and $\nu$ (section 3),
we have studied, in section 4, the TWS integrable sub-dynamics. 
The changes of the
$\cal P$ topology and, more specifically, the changes of the structure 
of orbit bundles filling $\cal P$ have been considered under the variations
of $\tau$ and $\nu$, and the ensuing effects on dynamics have been dicussed.  
In particular, the ballistic orbit bundles have been
found to have a periodic evolution whose time scale increases when the
orbit approaches the separatrix ${\cal S}_0$. The possibility of having
arbitrarily large time periods for a nonzero set of orbits disapperas
as soon as a coalescence process destroys ${\cal S}_0$. Various
situations in which the phase undergoes either a pulsed evolution or
a stepwise variation while $D/N$ shows rapid population inversions
have been discussed. These effects share the common feature of being
macroscopic, offer new information on the special TC regime of the
three-well system, and thus are interesting in order to design new 
experiments on coupled BECs. The analysis we have performed on the
$\cal P$ topology extends the study of the two-well dynamics~\cite{RAG}, 
and recognizes the presence of the two-well phenomenology within the 
TWS dynamics. We notice that similar phenomena have been observed in 
the coherent-mode dynamics of weakly excited condensates~\cite{YUK},
the motion equations of which exhibit in fact a rather 
similar nonlinear form.  

We emphasize the fact that, beyond the rich scenario of 
dynamical behaviors, our analysis 
clearly displays both the way to prime macroscopic effects and the
circumstances in which these are expected to occur by changing in a
controlled way the relevant parameters $N$, $\tau$ and $\nu$.
In section 5 we have presented a preliminary study of the purely 
quantum TWS. A nice result which confirms the utility of the new
canonical picture adopted in section 2 is that the algebraic analysis
of the microscopic degrees of freedom capable of perturbing
the integrable subregime permits to evidence the stable nature of 
TC configurations. These, in a sense, appear to be protected
by the energy gap ensuing from the discrete form of the energy
spectrum in the compact sector of the hamiltonian algebra su$(1,1)$.
Noticeably, the noncompact sector giving a continuous spectrum turns
out to characterize the system in correspondence to unstable TC 
states. A systematic quantum study of the TWS along the lines of 
Ref. \cite{FP}, which is presently in progress, will be presented 
elsewhere.

\bigskip

%
\centerline{ ACKNOWLEDGEMENTS}
\medskip

The financial support of the INFM (Italy) and of
the MURST is gratefully acknowledged.


\end{twocolumn}


\begin{thebibliography}{000}
\bibitem{com}
We remark that points such as maxima
$M_{\pm}$ in $\cal P$ might correspond to saddles in the full
phase space $\cal M$. This information can be evinced from the 
analysis of Ref. \cite{BFP1}. Here, the standard terminology
is used on $\cal P$ for the dimerlike dynamics, and serves to 
identify unambiguously extremal points of $H_0$.
\bibitem{COM1}
The stability analysis evidences that the trimer may have 
stable saddle points. These, however, never have a dimeric character. 
See ref. \cite{BFP1} for further detail
\end{thebibliography}
\end{document}